\documentclass[12pt,preprint]{aastex}

\slugcomment{}

\begin{document}

\title{ Limits on the AGN activities in X-ray underluminous galaxy groups }

\author{ K. S. Dwarakanath \& Biman B. Nath}
\affil{Raman Research Institute, Bangalore 560 080, India}
\email{dwaraka@rri.res.in, biman@rri.res.in}

\shorttitle{radio emission in high entropy groups}
\shortauthors{ Dwaraka \& Biman}

\begin{abstract}

We have observed four X-ray underluminous groups of galaxies using the Giant Meterwave
RadioTelescope. The groups NGC 524, 720, 3607, and 4697 are underluminous in relation
to the extrapolation of the L$_{x}$ - T relation from rich clusters and do not show
any evidence of current AGN activities that can account for such a departure. 
The GMRT observations carried out at low frequencies (235 and 610 MHz) were aimed at
detecting low surface brightness, steep-spectrum sources indicative of past AGN 
activities in these groups. No such radio emissions were detected in any of these
four groups. The corresponding upper limits on the total energy in relativistic
particles is $\sim$ 3 $\times$ 10$^{57}$ erg. This value is more than a factor of
100 less than that required to account for the decreased X-ray luminosities
(or, enhanced entropies) of these four groups in the AGN-heating scenario.
Alternatively, the AGN activity must have ceased $\sim$ 4 Gyr ago, allowing the
relativistic particles to diffuse out to such a large extent ($\sim$ 250 kpc) 
that their radio emission could have been undetected by the current observations.
If the latter scenario is correct, the ICM was pre-heated before the assembly
of galaxy clusters.
\end{abstract}


\section{INTRODUCTION}
Recent studies of the hot intracluster medium (ICM) in 
clusters of galaxies have shown that this gas has been heated by some
non-gravitational sources other than the heating expected from falling into
the gravitational potential of  dark matter. In particular, the
entropy of the gas, defined as $S\equiv T/n_e^{2/3}$, where T and n$_e$ denote
the gas temperature and the electron number density respectively, appears to be in excess
of expectations from gravitational
interactions of gas with dark matter. This excess entropy makes the
gas less luminous than expected and  affects a deviation
 from the scaling relation between X-ray luminosity 
and gas temperature, especially in poor clusters 
(Ponman et al. 2003, and references therein).
There have been a number of proposals to explain this entropy floor 
including energy input from
supernovae (e.g., Wu et al 2000), warm-hot
intergalactic medium (Valageas et al
2003), radiative cooling (e.g., Voit \& Bryan 2001), accretion shocks (e.g.,
Tozzi \& Norman 2001), and AGN heating (e.g., Nath \& Roychowdhury 2002).

One of the theoretical models to explain the X-ray observations invokes
energy input from active galactic nuclei (AGN).
It has been proposed that AGN heating not only can slow down or even balance 
the precipitous cooling of the gas in the central region, but it can 
also raise the gas entropy outside the core (r $>$ 100 kpc). 
These models were partly motivated by the presence of X-ray-deficient bubbles in
the inner regions of many clusters, for example, the Hydra A cluster 
(McNamara et al. 2000), or in the periphery of some clusters, for example,
in MKW3s (Mazzotta et al. 2002). 

Such bubbles, presumably fed by AGN activity,
 are expected to rise in the cluster atmosphere because of buoyancy
and as they expand in size, they are expected to do work against the ambient
gas, providing a source of heating for this gas.
Begelman (2001) and Ruszkowski \& Begelman (2002) suggested  this
`effervescent heating' could explain the related problem of quenching cooling
flows in galaxy clusters. Roychowdhury et al (2004, 2005) studied the
effect of `effervescent heating' outside the core and found that it could
explain the X-ray observations. According to their findings, a cluster with
a total mass $\sim 10^{14}$ M$_{\odot}$, with gas at $\sim 1.5$ keV,
 requires an AGN, or a number of 
AGNs, that could supply
energy of order $5 \times 10^{61}$ erg
over a time scale of $\sim 1$ Gyr (Roychowdhury et al 2005, Fig. 4).

Recently, Croston et al (2005) studied  the X-ray luminosity and temperature
relation in a sample of poor clusters after having divided
the sample into two groups : clusters with and without radio-loud objects in
 them. They found tentative evidence that clusters with radio-loud objects
in them deviated more from the scaling relation than clusters without
radio-loud objects (Fig. 1). 
This lent credence to the idea that AGNs are responsible
for the excess entropy in poor clusters.
Croston et al (2005) estimated the energy required to lower the 
entropy of a group to
sufficiently change its position in the X-ray luminosity 
and temperature plot, to be
$\ge 10^{58\hbox{--}59}$ erg.

However, if one extrapolated the cluster L$_{x}$-T relation,
it would fall above all data points,
indicating that all poor clusters in their sample deviated from the scaling
relation, although clusters with radio-loud objects in them deviated the
most (Figure 1). Such a deviation is clearly indicative of
factors other than only the gravitational interactions responsible
for the entropy of the ICM gas (Roychowdhury \&
Nath 2003). 

Theoretical modeling of the AGN heating by Roychowdhury et al (2004, 2005)
suggests that the gas entropy can remain high for a considerable period of time
($t \sim 10^{10}$ y)  even after the AGN activity has ceased. It is therefore 
reasonable to assume that the
reason for the increased entropy in some  galaxy groups can be the past 
activity of
an AGN in them, and not necessarily the current phase of activity. 
One way to probe the
past activity of an AGN is to study the low energy relativistic particles 
via synchrotron
radio emission at low frequencies ($<$ 1 GHz). The synchrotron life time of
relativistic
electrons with $\sim 10$ GeV, which can radiate at $\sim 600$ MHz for $B
\sim 1 \, \mu$G, is 
$\sim 1.4 (E/10 \, {\rm GeV})^{-1} (B/ 1 \, \mu {\rm G})^{-2}$
Gyr.

We selected four groups of galaxies, viz. NGC 524, NGC 720,
NGC 3607 and NGC 4697, from the sample of radio-quiet groups in Croston et al
(2005), which were among the most deviant groups from the L$_{x}$-T 
scaling relation (Fig. 1). 
In the context of  the AGN heating scenario, one way to understand the
deviant radio-quiet groups is to propose the existence of low surface brightness
and steep-spectrum radio emission in them which have not been 
detected by the existing surveys.
Amongst the 4 groups mentioned, 3 have not been observed at low ($\sim$ 300 MHz)
frequencies.  It therefore  appeared that 
these four groups (NGC 524, 720,
3607, 4697) were ideal targets for a deep radio continuum imaging at 
low frequencies with the Giant Meterwave Radio Telescope (GMRT).

\section{OBSERVATIONS}

The four groups NGC 524, NGC 720, NGC 3607 and NGC 4697 were observed at 
235 and 610 MHz
using the Giant Meterwave Radio Telescope. Some details of the observations 
are given
in Table 1. A 6 hour synthesis was carried out in each case, 
ensuring good visibility
coverage so as to enable detection of anticipated weak and 
extended radio emission. The visibility coverages of these observations
were adequate to detect sources up to an extent of $\sim$ 40$'$.
The data analysis was carried out using the Astronomical Image 
Processing System (AIPS).
On an average, $\sim$ 10 \% of the data was lost to radio frequency 
interference (RFI) at 235 MHz. The data lost to RFI was much less than this 
at 610 MHz.
The edited data was imaged after gain and bandpass calibration. The usual 
polyhedron
imaging was carried out (using standard tasks in AIPS) to account for the well-known effects of large 
fields of view
at low frequencies and the different heights of the antennas 
(the so-called 'w-term').
Several rounds of self-calibration and imaging were carried out to 
converge on the
best images. The RMS achieved in each case is given in Table 1. 
In all cases, the
RMS values in the images were dynamic range limited to $\sim$ 1 part in 
1000 of the peak flux density
in the respective images. As an illustration of the quality of the images 
obtained, 
images at 235 and 610 MHz of a small (but, the same) region of the sky near
the group NGC 4697 are shown in Figs. 2 \& 3. 

No diffuse radio emission was detected in any of the groups to a 
5$\sigma$ surface 
brightness limit of $\sim$ 10 mJy/beam at 235 MHz. The 5$\sigma$ surface brightness
detection limit at 610 MHz was $\sim$ 1 mJy/beam. These images made at a resolution
of $\sim$ 20$''$ were further smoothed to a resolution of $\sim$ 1$'$ in order
to detect extended features. 
The smoothed images at 235 MHz had an RMS of $\sim$ 12 mJy arcmin$^{-2}$.
No diffuse radio emission was detected in these smoothed images. 
 
\section{DISCUSSION}
We discuss the implications of our observations with regard to the energetics
of relativistic particles in the observed groups and the role of heating
of diffuse gas by relativistic particles from AGNs. Assuming synchrotron
emission at a given frequency $\nu$ 
from relativistic electrons emanating from a region of length scale
$l$, the minimum energy in relativistic particles (electrons and protons)
can be estimated from the standard minimum energy arguments (e.g., Longair
1994, p 292) as,
\begin{equation}
E_{rel} \approx 6 \times 10^{41} \, {\rm erg} \, \nu_8^{2/7} \, L_\nu^{4/7} \,
l_{kpc} ^{9/7} \, \eta^{4/7} \,,
\end{equation}
where $\nu_8=\nu/10^8$ MHz, $L_\nu$ is the power radiated by the source at $\nu$
(in W Hz$^{-1}$ sr$^{-1}$), $l_{kpc}$ is length scale in kpc and $\eta$
is the ratio of energy in protons to electrons. Converting $L_\nu$ to 
observed surface brightness, $I_\nu=L_\nu (V/l)$ (mJy arcmin$^{-2}$), where
$V\sim l^3$ is the source volume, this gives,
\begin{equation}
E_{rel}\approx 3.7 \times 10^{57} \, {\rm erg} \, \Bigl ({\nu \over 235 \,
{\rm MHz}} \Bigr )^{2/7} \, \Bigl ( { I_\nu \over {\rm mJY/arcmin}^2}
\bigr )^{4/7} \, \Bigl ({ l \over 100 \, {\rm kpc}} \Bigr )^{17/7} 
\, \Bigl ( { \eta \over 100 } \Bigr )^{4/7} \,.
\label{eq:min1}
\end{equation}
It is important to recall that the standard derivation integrates the
radio spectrum from a minimum to a maximum observed frequency,
typically from 10 MHz to 10 GHz.
It is also assumed that particles and magnetic fields fill the source
volume homogeneously and the field is assumed to be completely tangled
without any preferred direction.

If instead of using a fixed frequency interval one had used a fixed cut-off
in the energy of the relativistic particles, the results would be somewhat different. Beck
\& Krause (2004) re-derived the minimum energy/ equipartition energy argument
by assuming a relativistic proton spectrum which has a spectral index $\gamma
\sim 2.5$ down to energy $E_p \equiv m_p c^2$, and with a spectral index of
 zero below this (but above a low cutoff at, say, $E_1$). This assumption
was motivated by the observed spectrum of protons in our Galaxy. 
They assumed a fixed ratio between the number of relativistic
protons and electrons, of order $ K_0 \sim 100$, again motivated by observations
in our Galaxy at a few GeV. According to them, one has for for the
total energy density in relativistic particles,
at equipartition and for a spectral index of $\gamma =2.5$, 
\begin{equation}
\epsilon_{rel}\approx 4.8 \times 10^{-14}
\, {\rm erg/cc} \, \Bigl ({\nu \over 235 \,
{\rm MHz}} \Bigr )^{2/5} \, \Bigl ( { I_\nu \over {\rm mJY/arcmin}^2}
\bigr )^{8/15} \, \Bigl ({ l \over 100 \, {\rm kpc}} \Bigr )^{-8/15}
\, \Bigl ( { K_0 \over 100 } \Bigr )^{8/15} \,.
\end{equation}
This yields a total energy in relativistic particles ($E_{rel}=\epsilon
_{rel} l^3$), 
\begin{equation}
E_{rel}
\approx
1.4 \times 10^{57}
\, {\rm erg} \, \Bigl ({\nu \over 235 \,
{\rm MHz}} \Bigr )^{2/5} \, \Bigl ( { I_\nu \over {\rm mJY/arcmin}^2}
\bigr )^{8/15} \, \Bigl ({ l \over 100 \, {\rm kpc}} \Bigr )^{37/15}
\, \Bigl ( { K_0 \over 100 } \Bigr )^{8/15} \,.
\label{eq:min2}
\end{equation}
We find that these two estimates (from equations \ref{eq:min1} and 
\ref{eq:min2}) differ only by factors of order unity. We choose to use the
first equation to derive a conservative upper limit on the energy density
of relativistic particles based on the null results of our observations. 

To derive a limit, we need to find a suitable estimate of $l$, the extent
of the relativistic particles. In the absence of any other indicators, we
assume it to be comparable to the extent of the X-ray observations, namely,
the ROSAT extraction radius ($r_{cut}$, as described in Osmond \& Ponman
(2004)). The values of $r_{cut}$ of the respective groups are given in
Table 1. We find that for our sample, the
value of $r_{cut} \sim 60$ kpc.
The corresponding upper limit on the total energy of relativistic
particles for the four groups observed with GMRT is
therefore $3 \times 10^{57}$ erg (eq. \ref{eq:min1}). 
Since there can be uncertainties of order
unity stemming from a possible non-spherical geometry of the source, and other
considerations, we can put a firm upper limit on the total energy in
relativistic particles to be $\sim 10^{58}$ erg. 
Note that an r$_{cut}$ of 60 kpc
at the mean distance to these four groups corresponds to $\sim$ 10$'$ and 
well within the largest angular structures ($\sim$ 40$'$) up to 
which the current GMRT 
observations were sensitive.

The groups in our sample have gas temperature in the range of $0.3\hbox{--}
0.5$ keV, implying $M_{500} \sim 3.1\hbox{--} 7.7 \times 10^{12}$ M$_{\odot}$,
using the scaling relation from Finoguenov et al (2001).
This implies virial mass $M_{vir} \sim 10^{13}$ M$_{\odot}$, for a median value
of $M_{500} \sim 5 \times 10^{12}$ M$_{\odot}$, assuming a universal
profile of dark matter (Navarro, Frenk \& White 1997). It has been argued
that to reconcile with the X-ray observations of excess entropy in poor
clusters, it is required to inject energy of order $\sim 0.5\hbox{--}1$ keV
per particle (Cavaliere  et al 1999; Borgani et al 2002).
The number of particles in a $M_{vir} \sim 10^{13}$ M$_{\odot}$
cluster is $\sim (\Omega_b/\Omega_m) M_{vir} \sim 0.14  M_{vir}\sim 1.7\times
10^{69}$, and the total energy required to elevate its entropy to the observed
level is then estimated to be $\sim 1.5\hbox{--}3 \times 10^{60}$ erg.

Among the various possibilities of the basic mode of energy deposition,
AGN heating through `effervescent heating' has been recently studied in
some detail ---with bubbles of relativistic particles, that are created by
jets of AGN when they are active, and which detach from the host galaxy
 after the cessation of jet activity, and rise in the cluster atmosphere
because of buoyancy, expand and do pdV work against the ambient medium
(Begelman 2001; Roychowdhury et al 2004, 2005). Roychowdhury et al (2005)
calculated the required energy from AGNs, which are assumed to be active
for a period of $5 \times 10^8\hbox{--} 5 \times 10^9$ yr, after which the
ambient gas is allowed to evolve for a Hubble time at which point its
entropy (at two radii, $0.1 r_{200}$ and at $r_{500}$) is required to match
the observed values.
If the lowest envelope of Figure 4 in Roychowdhury et al (2005), 
representing the minimum energy required
from AGNs to enhance the entropy of clusters, is extrapolated down to
this mass range, then the energy required is $\sim 10^{60}$ erg.
There is,
therefore, a mismatch of a factor of $\sim$ 100 in the total energy estimate
and the limit from observation.

It could be argued that the observational limit is only a {\it lower
limit}, since it is based on equipartition, and is, therefore, close to the
minimum amount of energy that relativistic particles can harbor. In this
regard, we note that the energy estimate scales as $B^2$ for $B \gg B_{eq}$,
where $B_{eq}$ represents the equipartition magnetic field strength, and
 as $B^{-3/2}$ for $B \ll B_{eq}$. To reconcile with the above
 energy estimate, the magnetic field is required to be
either $B/B_{eq} \sim 10^{-4/3}$ or $B/B_{eq} \sim 10$ to reconcile with
the energy mismatch. The equipartition value, for an energy density
of $4 \sim 10^{-14}$ erg cm$^{-3}$ (eq. 3), 
is $B_{eq} \sim 1 \, \mu$G. Although there are 
uncertainties in the estimated strengths of
 magnetic fields
in clusters, it is difficult to presume that the strength 
would differ by an order of magnitude from this value,
as is required by the present observations (see also De Young 2006).
Moreover, although the factor $\eta$ could be somewhat larger than the fiducial
value of $\sim 100$ (Fabian et al 2002; De Young 
2006), with protons dominating the energy budget,
it is unlikely to change the energy estimate because of the weak dependence on
$\eta$.

Besides the strength of the magnetic field, there is an uncertainty about the
source size and the assumption that $l \sim r_{cut}$. 
The extent of the region occupied by
the relativistic particles can be estimated from diffusion length scales.
Observations of radio halos in clusters such as Coma suggest that
the diffusion coefficient of GeV particles in
a magnetic field of $B \sim 2\,\mu$G is $D \sim
(1\hbox{--}4)\times 10^{29}$ cm$^2$ s$^{-1}$ (Schlickeiser et al 1987).
The corresponding diffusion length scale in a Gyr, assuming a diffusion
coefficient of $\sim 2 \times 10^{29}$ cm$^2$ s$^{-1}$ is 
$r_D \sim \sqrt{(6 D t)} \sim 60  \,
(t/10^{9} \, {\rm yr})^{1/2}$ kpc.  
We therefore find that $r_D \sim 
 r_{cut} \sim 60 $ kpc
for the groups in our sample if the groups are 1 Gyr old. 
%Incidentally, the radius of clusters with mass $\sim 10^{14}$
%M$_{\odot}$ is $r_{200} \sim 0.8$ Mpc (at $z=0$). In other words, for the
%clusters in our sample, $r_cut \sim 0.1 r_{200}$
For a time scale of $t \sim 4 $Gyr, the particles would diffuse out to
a region of extent $\sim 2 r_{cut}$. 

As mentioned earlier, our observations are sensitive only up to an angular
size of  $\sim 40'$, which corresponds to a length scale (at the redshift
of our sample clusters) of $4 r_{cut}$. If the extent of the region occupied
by the relativistic particles is greater than $4 r_{cut}$ the current observations
would be insensitive to detect radio emission from such a region. Our observations can, therefore,
put a limit on the value of $l \ge 2 r_{cut}$, and in turn, put a limit
on the age of the relativistic particles as being larger than $\sim 4 $ Gyr. 
Our observation therefore puts a
limit on the epoch of any AGN activity as being at least  4 Gyr before the
present epoch. Incidentally, a recent proposal of heating the ICM with
cosmic rays from AGNs at $z \ge 2$
(which are also associated with the production
of Lithium-6 in halo metal poor stars) is consistent with this conclusion
(Nath, Madau, Silk 2006).

This time scale can also be compared with the epoch of formation of
structure with masses comparable to the total masses of the groups considered
here ($\sim$ 10$^{13}$ M$_{\odot}$) in the hierarchical structure formation scenario.
One can estimate the probability distribution
of the redshift of formation, defining it to be the epoch by which a fraction
of 3/4 of the total virial mass has been assembled from merger, given the
redshift of observation, that is, the present epoch.
In the $\Lambda$CDM cosmology, the duration available for evolution after
most of the mass has been assembled for a $10^{13}$ M$_{\odot}$ is estimated
to be however
$\sim 4 \times 10^9$ yr (Nath 2004). We can also find from 
Figure 1 of Voit et al (2003) that for such a structure, 3/4th of its
total mass was assembled at $\sim t/t_0 \sim 0.7$, where $t_0\sim 13.47$ Gyr
is the present age of the universe (Spergel et al 2003); the time available for further evolution
is therefore $\sim 4 $ Gyr. 

Our observations therefore put stringent limits on the model of AGN heating
in clusters for enhancing the entropy of the ICM. The null observations 
in the case of four groups with excess entropy
indicate that if AGN heating is the cause of heating the IGM, 
then it occurred well before the
mass assembly of these clusters--- in other words, the ICM was pre-heated
before the formation of the clusters.

\section{CONCLUSIONS}
The galaxy groups NGC 524, 720, 3607, and 4697 do not show any evidence of
either current, or past AGN activities to account for their decreased X-ray
luminosities corresponding to their temperatures. Any possible AGN contribution
to the heating of the ICM is a factor of $\sim$ 100 lower than that required to explain
the decreased X-ray luminosities of these groups. Alternatively, any possible AGN
activity must have occurred more than $\sim$ 4 Gyr ago for the current observations
to have not detected its radio emission. The latter scenario implies that the
ICM was pre-heated before the formation of the clusters.

\acknowledgements We thank Dr. Mitchell C. Begelman for his comments on the
manuscript.

\clearpage

\begin{deluxetable}{ccccc}
\tablenum{1}
\tablecolumns{5}
\tablewidth{0pc}
\tablecaption{GMRT Observations}
\tablehead{
\colhead{Source}&\colhead{R.A. (J2000)}&\colhead{Dec. (J2000)} 
&\colhead{RMS} & \colhead {r$_{cut}$}
\\
\colhead{}&\colhead{hh mm ss}&\colhead{o ~~$'$ ~~$''$} 
&\colhead{(mJy beam$^{-1}$)} & \colhead{(kpc)}
}
\startdata
NGC 524 & 01 24 47.7 & +09 32 20 & 3.8 & 56 \\
NGC 720 & 01 53 00.0 & -13 44 19 & 2.1 & 65 \\
NGC 3607 & 11 16 54.6 &+18 03 07 & 1.9 & 62 \\
NGC 4697 & 12 48 35.9 &-05 48 03 & 1.8 & 53 \\
\enddata
\tablecomments { Each of the sources were observed at 235 and 610 MHz for a total of 6 h.
For all the observations the synthesized beam is 20$''$ X 20$''$ at 235 MHz 
and 16$''$ X 16$''$ at 
610 MHz. The RMS quoted is at 235 MHz. The RMS at 610 MHz is $\sim$ 0.2 mJy/beam in all
cases. } 
\end{deluxetable}

\clearpage

\begin{figure}
\epsscale{0.7}
\plotone{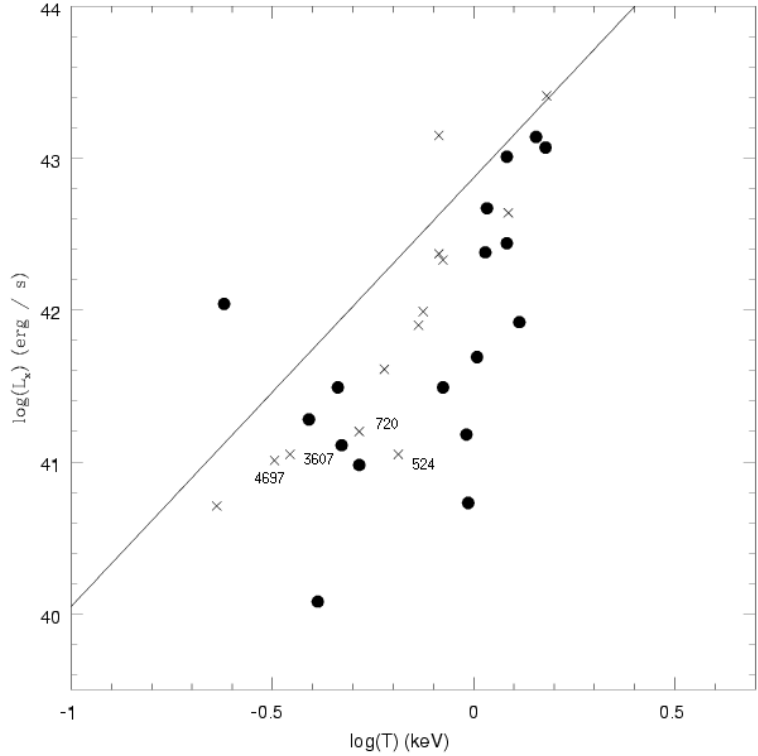}
\caption{X-ray luminosities of poor clusters and groups are shown
against their respective gas temperatures in keV. The data are from
Osmond \& Ponman (2004). The 'x' symbols are radio-quiet
groups and filled circles are radio-loud groups as in Croston et al 2004.  
The straight line is an extrapolation
of the L$_{x}$ - T  relation from rich clusters, as in, for example, Novicki 
et al 2002.
Note that even the radio-quiet
groups are below that estimated from this extrapolation.}
\end{figure}

\clearpage

\begin{figure}
\plotone{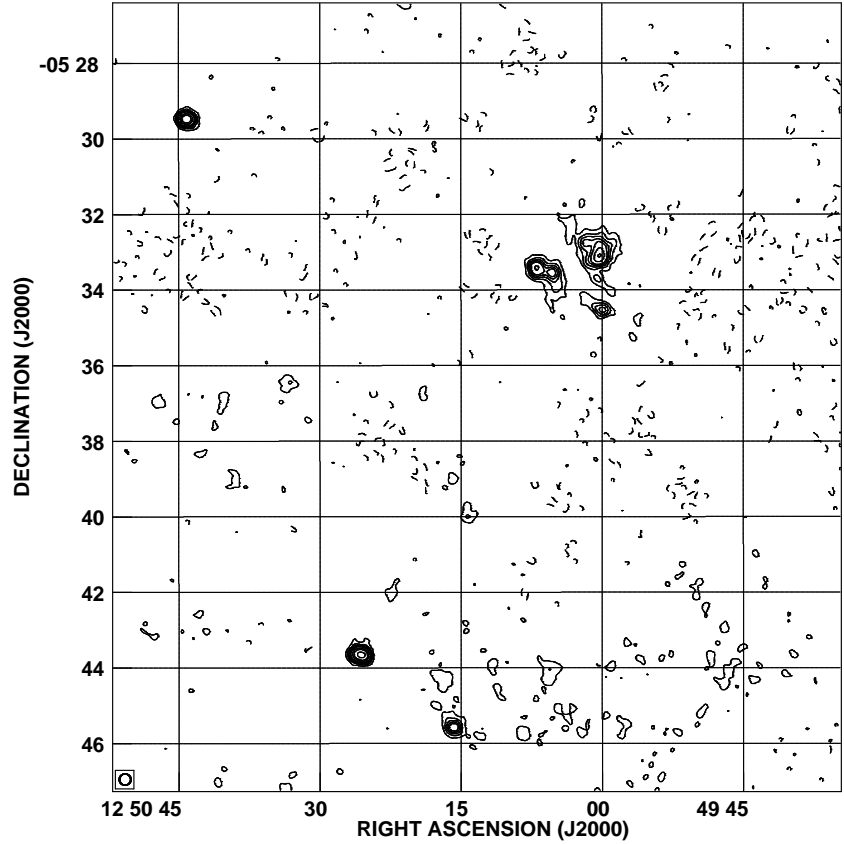}
\caption{ NGC 4697 : a small section of the radio continuum image at 235 MHz. 
The contour levels are at -4, -2, 2, 4, 6, 8, 10, 15, 20, 30, 40 and 50
times 2 mJy/beam. The synthesized beam is 18$''$ X 18$''$. The RMS on this 
image is 1.8 mJy/beam. The peak flux density in this field of view is 
1.3 Jy/beam.}
\end{figure}

\clearpage

\begin{figure}
\plotone{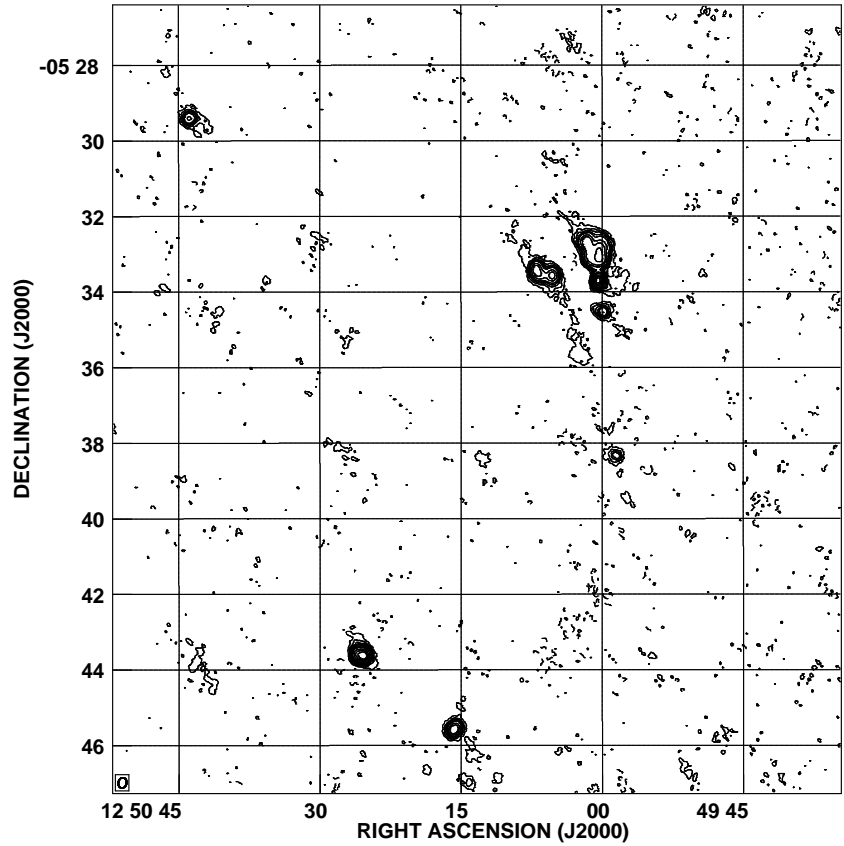}
\caption{ NGC 4697 : same region as in Fig. 2, but at 610 MHz. The contour
levels are at -4, -2, 2, 4, 6, 8, 10, 15, 20, 30, 40 and 50 times 0.2 mJy/beam.
The synthesized beam is 17$''$ X 13$''$ (PA = -11$^o$). The RMS on this image
is 0.15 mJy/beam. The peak flux density in this field of view is 0.23 Jy/beam.}
\end{figure}

\end{document}